\documentstyle[prd,twocolumn,dsfont,aps,amsmath,amsfonts,amssymb,epsfig]{revtex}

\draft
\begin{document}
\title {Estimating the post-measurement state}
\author{J\"urgen Audretsch${}^1$, Lajos Di\'osi${}^2$ and Thomas Konrad${}^3$}
\address{${}^{1,3}$Fachbereich Physik, Universit\"at Konstanz, Fach M674, 
D-78457 Konstanz, Germany \\ ${}^2$Research Institute for Particle and Nuclear 
Physics, H-1525 Budapest 114, POB 49.}


\maketitle

\begin{abstract}
We study generalized measurements (POVM measurements) on a single $d$-level
quantum system which is in a completely unknown pure state, and derive the
best estimate of the post-measurement state. The
mean post-measurement estimation fidelity of a generalized measurement is
obtained and related to the operation fidelity of the device. This illustrates
how the information gain about the post-measurement state and the
corresponding state disturbance are mutually dependent. The connection between
the best estimates of the pre- and post-measurement state is
established and interpreted. For pure generalized measurements the two
states coincide.
\end{abstract}


There are two important properties in which measurements on a
quantum system differ from measurements in classical physics:
Even if a finite number of identical copies of a system are
available, it is in general impossible to obtain complete
information about the state of the system. Furthermore, 
information can be extracted from a quantum system only at the cost 
of disturbing it.

These aspects are studied in the framework of quantum estimation
theory which has recently attracted much interest. It plays an
important role in quantum data processing in the context of
quantum information and computing. 
A typical topic is the determination of the
optimal fidelity of the estimated quantum state from 
$N$ identically prepared copies of the quantum system \cite{1}.
Algorithms for constructing an optimal positive operator valued
measure (POVM measurement) were discussed in \cite{2}. 
Adaptive projection measurements were treated in \cite{3}.
A related subject to the present discussion is the tension between
information gain and disturbance \cite{4}.
The balance between the mean operation fidelity and the
estimation fidelity of the pre-measurement state has been
studied by Banaszek \cite{5}. We will come back to his results later.

The purpose of this Letter is studying the estimation of the 
post-measurement state. Suppose a generalized measurement 
(POVM measurement) is performed on a single $d$-level system of 
pure but otherwise completely unknown quantum state.
Knowing the measurement result and the specifications of the measurement,
what is the best estimate of the post-measurement state and what is 
the corresponding highest fidelity? Of all measurements granting a certain
estimation fidelity, which is the one with the lowest disturbance?
And finally, how are the best estimations of the pre- and post-measurement 
states related? All these questions will be answered below in closed 
analytical forms.

Situations in which it is important to guess the ''post-measurement state'' 
are known from everyday life. Medical inspections with x-rays, radioactive 
chemicals etc. are invasive measurements as quantum measurements in general 
are. The more information such inspections provide the more damage they cause. 
No copy of the patient is available. The patient's state is therefore to be
estimated on the basis of a single run inspection whereby the
doctor has to decide about the strength of his intervention 
in choosing a balance between information gain and disturbance. 
Since, furthermore, any subsequent medical treatment must take into account
that an unavoidable disturbance has happened, it has to be adjusted to the 
post- and not to the pre-inspection state.

There are quantum informatic setups exhibiting such characteristic
traits. A typical example is a sequence of generalized 
measurements aiming at the monitoring of the state evolution of a single
quantum system \cite{6}. An important strategy to improve
the information and to diminish the disturbance is to
adjust the parameters of each forthcoming generalized measurement to the expected 
pre-measurement state \cite{7}. To this end, the post-measurement
state of the previous measurement must be estimated.
We will not work out this example, but rather turn to the
post-measurement state in general.

A given generalized measurement is described by a set of $n$
operators $M_s$,  where the index $s=1,\ldots ,n$ labels the
possible readouts of the measurement. These measurement operators, 
also called Kraus-operators, act on the quantum state of the measured 
system. One may think of a $d$-level system. The readout $s$ will in 
general not correspond to one of these levels, contrary to typical
projective measurements.
The pure pre-measurement state $|\psi \rangle$ of the system is changed by a 
generalized measurement with outcome $s$ into the conditional 
post-measurement state
\begin{equation}\label{psi'}
|\psi^{(s)} \rangle = \frac{M_s |\psi\rangle}{\sqrt{\langle
\psi | M_s^\dagger M_s |\psi\rangle}} \,\,.
\end{equation}
Obviously, $|\psi^{(s)} \rangle$ will always depend on the initial state
$|\psi\rangle$ unless the rank of $M_s$ is $1$. Therefore the post-measurement 
state remains in general unknown if $|\psi\rangle$ is unknown,
and can only be estimated. 
The probability for the measurement result $s$ to occur is given by
\begin{equation}\label{eq1:ps}
p_s = \langle \psi |E_s |\psi \rangle \,,
\end{equation}
where the operators $E_s$ are defined by
\begin{equation}
\label{defEs}
E_s := M_s^\dagger M_s \,\,.
\end{equation}
They are positive operators  satisfying a completeness relation
$\sum_{s=1}^n E_s = \mathds{1}$ which guarantees 
$\sum_{s=1}^n p_s = 1$ for the probabilities. The set $\{E_s\}$ is called a
positive-operator-valued measure (POVM) and the individual
operators $E_s$ are also known as POVM elements or effects.

To prepare later calculations we introduce the spectral
decomposition of $E_s$ 
\begin{equation} \label{Esrr}
E_s=\sum_{i=1}^d a_i^{(s)} |r_i^{(s)} \rangle \langle
r_i^{(s)} | \,\,.
\end{equation}
$a^{(s)}_i$ are the positive eigenvalues. The eigenvectors 
$\{|r_i^{(s)}\rangle\}$ form an orthonormal basis. 
Due to the polar decomposition theorem (cf. e.g.\cite{8}), we may split
the measurement operator $M_s$ into a product of a
unitary operator $U_s$ and the square root of $E_s$
\begin{equation}\label{Ms}
M_s = U_s \sqrt{E_s}  \,\,.
\end{equation}
This implies
\begin{equation}
M_sM_s^\dagger = U_sE_sU_s^\dagger \,.
\end{equation}
Thus the positive operators $M_sM_s^\dagger$ and $E_s$ have the same eigenvalues
$a_i^{(s)}$ and the diagonal representation of $M_sM_s^\dagger$ becomes
\begin{equation}\label{MsMs}
M_s M_s^\dagger = \sum_{i=1}^d a_i^{(s)} |l_i^{(s)} \rangle
\langle l_i^{(s)} | \,\,.
\end{equation}
The eigenvectors $|l_i^{(s)}\rangle=U_s|r_i^{(s)} \rangle$ form
again an orthonormal basis. Herewith and with the help of eqs.(\ref{Esrr}) 
and (\ref{Ms}) we obtain as result the useful 
{\em bi-orthogonal expansions} of the unitary operators $U_s$ and of the 
measurement operators $M_s$:
\begin{equation}\label{Uslr}
U_s=\sum_{i=1}^d |l_i^{(s)} \rangle \langle r_i^{(s)} | 
\end{equation}
\begin{equation} \label{Mslr}
M_s = \sum_{i=1}^d \sqrt{a_i^{(s)}} |l_i^{(s)} \rangle \langle
r_i^{(s)} | \,\,.
\end{equation}
$|l_i^{(s)}\rangle$ and $|r_i^{(s)}\rangle$ are the l.h.s.
and r.h.s. eigenvectors of $M_s$, respectively. The number of non-zero 
eigenvalues $\sqrt{a_i^{(s)}}$ equals the rank of $M_s$.

Based on this we can now move to the problems of quantum state estimation.
We assume a {\it single} $d$-level quantum system prepared in a completely
unknown pure pre-measurement state $|\psi \rangle$. 
A particular generalized measurement specified by the 
known set $\{M_s\}$ of operators is performed with measurement result $s$ 
which is read off. What is the optimal strategy for the estimation of the 
post-measurement state $|\psi^{(s)}\rangle$ prepared by the measurement?
It is worthwhile to emphasize that the only data available for the
estimation are the set $\{M_s\}$ specifying the measurement and the value
$s$ of the actual readout.

If the state $\vert \chi^{(s)} \rangle$ is proposed as an estimate of 
the unknown post-measurement state $|\psi^{(s)} \rangle$, the fidelity
\begin{equation} \label{eq:5}
f_s = |\langle \chi^{(s)}|\psi^{(s)} \rangle |^2 =
\frac{1}{p_s} |\langle \chi^{(s)} |M_s | \psi \rangle |^2
\end{equation}
is a measure of the quality of the estimation. The fidelity
$\overline f$ averaged over all measurement outcomes reads:
$\overline f = \sum_{s=1}^n  f_s  p_s$. 
The mean estimation fidelity $G_{\rm{post}}(\chi)$ 
in case the in-going (pre-measurement) state is {\em completely unknown}, 
is the result of an integration over all possible states $|\psi \rangle$:
\begin{equation}\label{H:=}
G_{\rm{post}}(\chi):= \int \overline f d \psi = \int d \psi \sum_{s=1}^n 
\langle\chi^{(s)} |M_s| \psi \rangle \langle \psi | M_s^\dagger
|\chi^{(s)} \rangle \,\,,
\end{equation}
with respect to the normalized unitary invariant measure on the state space,
yielding: 
\begin{equation}\label{H}
G_{\rm{post}}(\chi) = {\frac{1}{d}} \sum_{s=1}^n  \langle \chi^{(s)} |M_s
M_s^\dagger |\chi^{(s)} \rangle \,\,.
\end{equation}

By virtue of eq.(\ref{MsMs}), each component in the sum over $s$ in
(\ref{H}) is maximized if $|\chi^{(s)}\rangle $ is chosen to be the
eigenvector $|l_{\max}^{(s)}\rangle $ of $M_sM_s^\dagger$ of the
maximum eigenvalue $a_{\max}^{(s)}$. For the measurement result $s$ the 
{\em best estimate of the post-measurement state} is therefore given by
\begin{equation}\label{chi}
|\chi_{\rm{post}}^{(s)}\rangle = |l_{\max}^{(s)}\rangle.
\end{equation}
In case of degeneracy of the greatest eigenvalue $a_{\max}^{(s)}$,
any state vector from the corresponding eigenspace represents an
optimal estimation of the post-measurement state. 
The maximum value of $G_{\rm{post}}(\chi)$ reads
\begin{equation}\label{Hopt2}
G_{\rm{post}} = \frac{1}{d} \sum_{s=1}^n a_{\max}^{(s)}\,.
\end{equation}
$G_{\rm{post}}$ is the {\em mean post-measurement estimation fidelity}. 
$|\chi_{\textrm{post}}^{(s)} \rangle$ and $G_{\rm{post}}$
are determined solely by the operators $M_s$ which specify the generalized 
measurement.

We now address the question, how  $G_{\rm{post}}$ is related to the 
{\em mean operation fidelity} $F$ which
describes how much the state after the measurement resembles the
original one. The larger the value $F$ of a measurement is, the weaker is its
disturbing influence. Arguing as above, $F$ is obtained from eq.(\ref{H:=})
if we replace $|\chi^{(s)}\rangle$ by $|\psi\rangle$:
\begin{equation}
F=\int d \psi \sum_{s=1}^n |\langle \psi| M_s |\psi \rangle |^2\,\,.
\end{equation}
It may be rewritten as \cite{5}:
\begin{equation}
F=\frac{1}{d(d+1)}\Bigl(d+\sum_{s=1}^n|\textrm{tr}M_s|^2\Bigr)\,.
\end{equation}

To derive a relation between $G_{\rm{post}}$ and $F$, it is useful to first 
relate $G_{\rm{post}}$ to the estimation fidelity of the pre-measurement 
state. Denoting this estimate by $| \chi^{(s)}\rangle$, the corresponding mean 
estimation fidelity, in analogy to $G_{\rm{post}}(\chi)$ of eq.(\ref{H:=}), reads:
\begin{equation}
G_{\rm{pre}}(\chi)=\int d \psi \sum_{s=1}^n p_s |\langle \chi^{(s)} | \psi \rangle |^2
\end{equation}
which may be rewritten according to Banaszek \cite{5} as
\begin{equation}\label{G=frac}
G_{\rm{pre}}(\chi)= \frac{1}{d(d+1)} \left( d + \sum_{s=1}^n \langle \chi^{(s)}
|E_s |\chi^{(s)} \rangle \right) \,\,.
\end{equation}

The optimum pre- and post-measurement fidelities are closely related.
For a given measurement result $s$, the best estimate 
$|\chi_{\rm{pre}}^{(s)} \rangle $ of the
pre-measurement state is the one which maximizes the corresponding component
in the sum in eq.(\ref{G=frac}). Because of eq.(\ref{Esrr}), it is given by the 
eigenvector $|r_{\max}^{(s)} \rangle$ of $E_s$ belonging to the maximum 
eigenvalue \cite{5,9}. But this eigenvalue is again 
$a_{\max}^{(s)}$. The {\em best estimate of the pre-measurement state} 
related to the outcome $s$ is therefore
\begin{equation}\label{varphiins}
|\chi_{\rm{pre}}^{(s)} \rangle = |r_{\max}^{(s)} \rangle \,\,.
\end{equation}
We denote the corresponding maximum value of $G_{\rm{pre}}(\chi)$ by
$G_{\rm{pre}}$ and call it the {\em mean pre-measurement estimation fidelity}. 
Comparing it to the form (\ref{Hopt2}) of $G_{\rm{post}}$, we obtain the
simple new relationship:
\begin{equation}\label{eqGopt}
G_{\rm{pre}} = \frac{1}{d+1} (1+G_{\rm{post}}) \,\,.
\end{equation}

This result allows us to transcribe Banaszek's constraint \cite{5} between
$F$ and $G_{\rm{pre}}$ into a constraint relating $F$ and $G_{\rm{post}}$: 
\begin{equation}\label{sqrtF}
\sqrt{(d+1)F-1} \leq \sqrt{G_{\rm{post}}}+\sqrt{(d-1)(1-G_{\rm{post}})} \,\,.
\end{equation}

\begin{figure}[htb]
\centering \epsfig{file=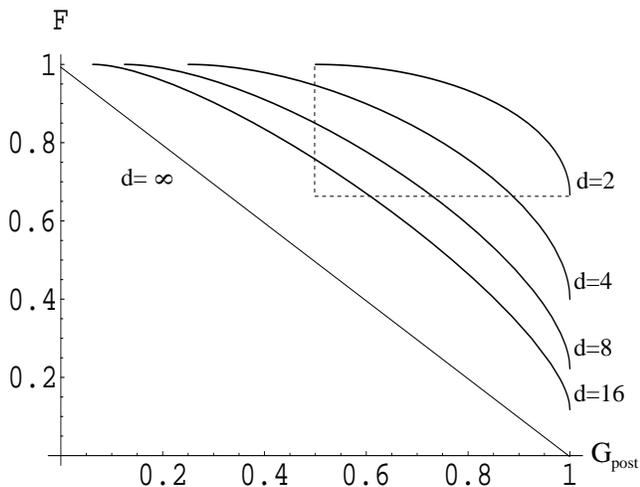,width=\linewidth}
\caption{Maximal operation fidelity $F$ for given estimation
fidelity $G_{\rm{post}}$ of the post-measurement state  in dimensions
$d=2,4,8,16, \infty$.The dashed lines mark for dimension $d=2$ the domain 
for possible combinations of $F$ and $G_{\rm{post}}$.}\label{fig:1}
\end{figure}

To illustrate how state disturbance and information gain are related for the
post-measurement situation, we display the domain of possible combination of 
$F$ and $G_{\rm{post}}$ in the $G_{\rm{post}}$-$F$ plane. 
If the system is not influenced at
all, the measurement has the operation fidelity $F=1$. In this case the guess
of the pre- and post-measurement state is totally random which amounts to
$G_{\rm{pre}}=G_{\rm{post}}=1/d$. On the other hand there are measurements
which allow to predict the post-measurement state exactly (e.g. projection
measurements), i.e., with maximum fidelity $G_{\rm{post}}=1$. 
This leads via eq.(\ref{eqGopt}) to $G_{\rm{pre}}=2/(d+1)$. 
This result for $G_{\rm{pre}}$ has also been obtained in \cite{1,4,5,6}. 
It is known \cite{2} that it corresponds to $F=2/(d+1)$. To summarize, the domain 
of possible combinations $(G_{\rm{post}},F)$ is limited by 
$1/d \le G_{\rm{post}} \le 1$ and $2/(d+1) \le F \le 1$ as well as by the 
inequality (\ref{sqrtF}). The boundaries of the domain are indicated in Fig. 
\ref{fig:1} for $d=2$, including the dashed lines. In this domain,
every particular generalized measurement $\{M_s\}$
corresponds to a point. Its position illustrates to what extent 
the information about the outgoing (post-measurement) state is gained at the cost of
disturbing the in-going (pre-measurement) one. 
Large values of $F$ combined with large values
of $G_{\rm{post}}$ characterize the most optimal type of generalized 
measurement. For increasing dimension $d$ of the state space all types of 
measurements become less advantageous (cf. Fig. \ref{fig:1}).

To complete this discussion we return to the question: What type of 
generalized measurements apart from projection measurements make it possible
to know the post-measurement state $|\psi^{(s)} \rangle$ exactly? 
As we mentioned earlier, the necessary condition is the rank of Kraus-operator
$M_s$ be $1$:
\begin{equation}\label{Ms1}
M_s = \sqrt{a^{(s)}} |l^{(s)} \rangle \langle r^{(s)} | \,\,.
\end{equation}
From eq.(\ref{psi'}) it follows that the post-measurement state is always
$|l^{(s)}\rangle$ independently of the otherwise unknown pre-measurement
state. If we apply our general rule (\ref{chi}) to this trivial case we find 
that, indeed, the best estimate is the true one: 
$|\chi_{\rm post}^{(s)}\rangle=|l^{(s)}\rangle$. The rule (\ref{varphiins})
yields $|\chi_{\rm pre}^{(s)}\rangle=|r^{(s)}\rangle$ for the best estimate
of the pre-measurement state. Hence the ultimate form of the rank-one 
Kraus-operators is this: 
\begin{equation}\label{Ms1'}
M_s = \sqrt{a^{(s)}} |\chi_{\rm{post}}^{(s)} \rangle \langle \chi_{\rm{pre}}^{(s)}|
\,\,.
\end{equation}
The corresponding effects $E_s$ are then given by
\begin{equation}\label{Es}
E_s = a^{(s)} |\chi_{\rm{pre}}^{(s)} \rangle \langle
\chi_{\rm{pre}}^{(s)}|\,\,.
\end{equation}
The completeness relation $\sum E_s=\mathds{1}$ constrains the 
pre-measurement state estimates to form an overcomplet basis in general.
The set of post-measurement states is not constrained at all.
Note that the multiplicity of different measurement results $s$ may 
exceed the number $d$ of levels in our system. Since neither  
$|\chi_{\rm{pre}}^{(s)}\rangle$ nor $|\chi_{\rm{post}}^{(s)}\rangle$  
have to form orthogonal systems they are in general not the eigenstates 
of any Hermitian observable. So we are still having a generalized 
measurement and not a projective measurement. 
The post-measurement state is nevertheless exactly known
$(G_{\rm{post}}=1)$ and the optimal estimate of the pre-measurement
state $|\chi_{\rm{pre}}^{(s)} \rangle$ is of maximal estimation fidelity
$G_{\rm{pre}}=2/(d+1)$.

We turn to a further aspect of information gain and state disturbance.
In eq.(\ref{Ms}) we have uniquely decomposed the measurement operation $M_s$
which corresponds to the measurement result $s$, into the positive operator
$\sqrt{E_s}$ and a unitary operator $U_s$. The unitary part does not change 
the von Neumann entropy. By virtue of eq.(\ref{eq1:ps}), 
all information, which is contained in a measurement result, 
goes back to $\sqrt{E_s}$. In particular, the estimation fidelities 
$G_{\rm{post}}$ and $G_{\rm{pre}}$ (\ref{Hopt2},\ref{G=frac}) depend 
only on the eigenvalues of $E_s$. The part $\sqrt{E_s}$ of $M_s$ represents,
at a given information gain, the unavoidable minimal disturbance of the state 
vector. We call $\sqrt{E_s}$ the {\em pure measurement part} of a generalized 
measurement and a measurement with $U_s=\mathds{1}$ a 
{\em pure measurement}. 
The operation fidelity $F$ depends on the unitary parts $U_s$, too. The inequality
(\ref{sqrtF}) shows that the maximal operation fidelity $F$ is limited by
$G_{\rm{post}}$ and therefore by the pure measurement part.

Having connected the three mean fidelities $F, G_{\rm{pre}}$ and 
$G_{\rm{post}}$, we
connect now the best guesses $|\chi_{\rm{pre}}^{(s)}\rangle$ and
$|\chi_{\rm{post}}^{(s)} \rangle$ for the pre- and
post-measurement states, respectively. 
The two best guesses are the distinguished pair of l.h.s. and r.h.s. 
eigenvectors to the same eigenvalue, cf. eqs.(\ref{chi}) and 
(\ref{varphiins}). Invoking the expansions 
(\ref{Uslr}) and (\ref{Mslr}), this leads directly to the results
\begin{equation}\label{Usvarphi}
U_s |\chi_{\rm{pre}}^{(s)} \rangle = |\chi_{\rm post}^{(s)} \rangle
\end{equation}
and
\begin{equation}\label{fracMs}
\frac{M_s |\chi_{\rm pre}^{(s)} \rangle}{\sqrt{a_{\max}^{(s)}}}
= |\chi_{\rm post}^{(s)} \rangle \,\,.
\end{equation}
Equation (\ref{Usvarphi}) shows that the best estimate for the
post-measurement state can be obtained from the best estimate of the 
pre-measurement state by applying merely the unitary part $U_s$ of the 
measurement operator. This has the surprising consequence that for all
pure measurements the best estimations for the pre- and
post-measurement state always agree if the in-going state $|\psi \rangle$ is
completely unknown. This is the case regardless
of the values of the operation fidelity $F$ and the estimation
fidelities $G_{\rm{pre}}$ and $G_{\rm{post}}$. 

Finally we give a physical interpretation of relation (\ref{fracMs}). 
As a matter of fact, both the pre- and post-measurement states become 
only partially revealed by the estimation procedure. Nonetheless,
even non-optimal estimates $|\chi_{\rm pre}^{(s)}\rangle$, 
$|\chi_{\rm post}^{(s)}\rangle$ must obey the constraint (\ref{fracMs})
expressing the certain fact that the post-measurement state results from
the generalized measurement (\ref{psi'}) of the pre-measurement one.  
Recall that we estimated the optimum pre- and post-measurement states by 
maximizing independently the pre- and post-measurement fidelities. 
We did not guarantee explicitly that the two optimum states 
$|\chi_{\rm pre}^{(s)}\rangle$,
$|\chi_{\rm post}^{(s)}\rangle$ satisfy the exact constraint.
The derived result (\ref{fracMs}) proves that they do.

In conclusion, we have studied generalized measurement $\{M_s\}$ on a 
single $d$-level quantum system. For the case when the initial state is 
pure and otherwise completely unknown, we pointed out that
the best estimates of the pre- and post-measurement states for a given 
measurement readout $s$ are the respective right and left eigenvectors
of $M_s$, belonging to the (common) largest eigenvalue. 
The mean post-measurement estimation fidelity of the 
measurement device is also calculated and shown to satisfy a simple
relationship with the mean pre-measurement estimation fidelity. 
A constraint between the post-measurement estimation fidelity and the 
operation fidelity of the measurement illustrates how state disturbance and
information gain about the post-measurement state are competing with each other. 
We have shown that for pure generalized measurements the independent best 
estimates of the pre- and post-measurement states agree. We have proved that, 
in general, they are related via the corresponding measurement operator as 
we expect of them.

This work was supported by the Optik Zentrum Konstanz.
L.D. acknowledges the support from the Hungarian Science Research Fund
under Grant 32640.

\end{document}